\documentclass[a4paper,10pt,pra,twocolumn]{revtex4}
\usepackage[dvips]{graphics,color}
\usepackage{amsmath}
\usepackage{amsfonts}
\newcommand{\one}{\mathrm{I} \! \! 1}
\usepackage{amssymb}
\usepackage{amstext}
\begin{document}

%%%%%%%%%%%%%%%%%%%%%%%%%%%%%%%%%%%%%%%%%%%%%%%%%%%%%%%%%%%%%%%%%%%%%%%%%%%%%%%
\title{Classicality of spin coherent states via entanglement and distinguishability}
\author{D. Markham and V. Vedral}
\affiliation{Optics Section, Blackett Laboratory, Imperial
College, London SW7 2BW, United Kingdom.}

%%%%%%%%%%%%%%%%%%%%%%%%%%%%%%%%%%%%%%%%%%%%%%%%%%%%%%%%%%%%%%%%%%%%%%%%%%%%%%%
\begin{abstract}
We trace the resistance to entanglement generation of spin
coherent states when passed through a beam splitter as we vary $S$
through $S=1/2\rightarrow\infty$. In the infinite $S$ limit the
spin coherent states are equivalent to the high-amplitude limit of
the optical coherent states. These states generate no entanglement
and are completely distinguishable. This transition is discussed
in terms of the classicality of the states. The decline of the
generated entanglement, and in this sense increase in classicality
with $S$, is very slow and dependent on the amplitude $z$ of the
state. Surprisingly we find that, for $|z|
>1$, there is an initial increase in entanglement followed by an
extremely gradual decline to zero. Other aspects of classicality
are also discussed over the transition in $S$, including the
distinguishability, which decreases quickly and monotonically. We
illustrate the distinguishability of spin-coherent states using
the representation of Majorana.
\end{abstract}
%%%%%%%%%%%%%%%%%%%%%%%%%%%%%%%%%%%%%%%%%%%%%%%%%%%%%%%%%%%%%%%%%%%%%%%%%%%%%%
%
%\date{\today}
\maketitle

\section{Introduction} \label{sc:into}
The transition from the quantum to the classical world is not
fully understood at present. Although classical physics is
believed to be a limiting case of quantum mechanics, it is
frequently unclear which limit should be taken and how this is to
be achieved. We can, for example, try to think of how the laws of
physics come to appear as the classical laws if we start from the
Schr\"odinger equation, as in \cite{Gell-Mann92}. Alternatively,
we might think about how physical objects themselves come to
appear classical, when they are so manifestly quantum at the most
basic level that we can test in our most accurate experiments. In
this case, initial quantum states describing objects would somehow
have to make a transition to become classical. Quantum physics has
been considered to approach the classical in many different ways
and we name several of them. (1) As $\hbar$ goes to zero, the
variance (or uncertainty) of two non commuting measures go to zero
and we can separate all states perfectly (this is known as Bohr's
correspondence principle); also, the classical Hamilton-Jacobi
dynamics can be ``derived" from the Schr\"odinger equation in this
limit. (2) Positivity of the Wigner or $P$ function is often taken
to imply classicality, since they then can be thought of as
representing real probability functions as in classical
statistical mechanics. (3) The entanglement present in a
collection of states is clearly an indication of nonclassicality
and a lack of it can be construed as a signature of classicality
(various decoherence pictures, on the other hand, use the
entanglement with the environment as an indication of
classicality; in this case the global presence of entanglement
leads to classicality). (4) Distinguishability of states
(classical states are always fully distinguishable at least in
principle). Although each of these criteria is adequate in its own
domain, it needs to be stressed that they are by no means
equivalent in general. In fact, they frequently contradict each
other. A well known example of this is that a state of two light
modes can be entangled and still have an overall positive Wigner
function, as well as vice versa. It is, therefore, very important
to investigate the relationship between these various approaches
in order to gain a deeper understanding of the transition between
the quantum and the classical. In this paper we trace the
spin-coherent state through a range of spin $S$, from $S=1/2$ to
the limit $S\rightarrow \infty$, and comment on this as a
transition to classicality for this class of pure states.

One of the first set of states defined specifically with
classicality in mind is the optical coherent states or Glauber
states \cite{Glauber63}. These are minimum-uncertainty states and
their creation can be implemented with classical currents. The
Glauber states can then be generalised in different ways to a
wider set of states, including minimum-uncertainty-defined states
and group-defined states (see \cite{Klauder} and \cite{Klauder01}
for good reviews). One such generalisation is the set of spin
coherent states, introduced in \cite{Radcliffe71} and
\cite{Arecchi72}, which have been compared to the Glauber states
and some parallels drawn. For example, they too are states of
minimum uncertainty and they can be produced by classical fields
acting on the ground state. Historically, one claim of
classicality of the Glauber states is that they are the only pure
states to remain unentangled when passed through a beam splitter
\cite{Aharonov66} (which relates to one aspect of classicality
mentioned above). More recently it has been shown that a mixture
of these states also remains unentangled when passed through a
beam splitter \cite{Kim02,Xiang-bin02} (implying that generation
of entanglement requires ``nonclassical" states). Unlike the
Glauber state however, the spin-coherent state can produce
entanglement when passed through a beam splitter, even a maximally
entangled state for the spin $1/2$ case. In the $s \rightarrow
\infty$ limit though, the spin-coherent states go to the high
amplitude limit of the Glauber states
\cite{Radcliffe71,Arecchi72}, and hence the entanglement goes to
zero.

In this paper, we extend this comparison of the spin-coherent
states to looking at the transition as we change the size of the
spin from $S=1/2$ to the limit $S\rightarrow \infty$ Glauber
states. This is done with respect to their robustness against the
entanglement created through a beam splitter. As well as being
interesting in its own right, this property allows us to
investigate how quickly the classical features of the Glauber
states mentioned above appear. We discuss the transition as an
approach to this classicality, considering the ``most classical"
to be the limiting Glauber states. We note, however, that this is
of course by no means a complete study of classicality. We deal
here only with pure states, and classically we can only truly
model mixed states. Indeed, even then the states are, of course,
still quantum and they behave like classical states only if the
quantum features are small enough to be ignored \cite{Vogel00}. In
this paper we ignore the dynamical aspects. Even when states are
``classical", one could argue that their dynamics may not be
(although, with respect to this, it should be pointed out that
under the influence of a classical electric field - essentially a
rotation - the spin-coherent states remain spin coherent states,
as proved by Arecchi {\it{et al}}.\cite{Arecchi72}); for example,
the complexity of quantum dynamics may be very different from that
of classical. We talk about classicality in a very specific sense,
namely, in entanglement and distinguishability, for which, in the
specific scenarios discussed, the Glauber state is the most
classical among the set of pure spin-coherent states. In this
sense, we trace the spin-coherent state from what we think of as
the ``least" classical, $S=1/2$, to the ``most" classical $S
\rightarrow \infty$.

In particular, in Sec. \ref{sc:BS} we investigate the entanglement
of the state generated by passing it through one arm of a 50:50
beam splitter while the other arm is left in the empty vacuum
state. We present a different proof to those of \cite{Radcliffe71}
and \cite{Arecchi72} that the spin-coherent states tend to the
Glauber states in the infinite $S$ limit, in terms of the states
in the first quantisation. Our method has the advantage that it
easily follows that there are infinitely many other states of the
same dimensionality as spin-coherent state with the same property
that they asymptotically approach the Glauber states, which is not
imediately obvious from \cite{Radcliffe71} and \cite{Arecchi72}.
We find that the reluctance to generate entanglement is very slow
to increase with $S$ and the entanglement can be near zero only
for very high $S$. In addition, the entanglement generation is
dependent on $z$ and for $|z| > 1$ we see a surprising rise in
entanglement with $S$- hence, a decline in classicality in this
sense. This is explained and parallels drawn to another similar
case noted by Arnesen {\it{et al}}. \cite{Arnesen01}.

In Sec. \ref{sc:disc}, we discuss this transition in terms of
classicality as viewed as a reluctance to create of quantum
correlations (entanglement). We then discuss different ways in
which this transition can be viewed in terms of classicality. We
focus on one classical feature, that of distinguishability, and
illustrate the changes with $S$ using the Majorana representation
\cite{Majorana32} via the Helstrom optimal measurement
\cite{Helstrom,Fuchsthesis} (which gives different measures and
success probabilities for different states, depending on how close
they are to orthogonal). For a $d$-dimensional state, the Majorana
representation defines the state (up to a global phase) by $d-1$
points on the surface of a Riemann sphere. Geometric
interpretation of physical systems is often very helpful in
gaining further intuition. For example, the Bloch sphere can be
used to see intuitively how optimal fidelity is achieved in
universal quantum cloning \cite{Gisin98}. The Majorana
representation shows well the changes in the measures affected by
changing $S$ for spin-coherent states in a simple geometry.

\section{Action Through A Beam Splitter} \label{sc:BS}

In quantum optics the action of a beam splitter is described as a
particular kind of mixer of two beams of light or, more
mathematically, two modes of the quantised electromagnetic field.
A beam splitter takes an input state comprised of a general
$n$-photon Fock state in one mode and a ground (vacuum) Fock state
in the other, $|n,0\rangle$, to a superposition of binomially
populated modes:
\begin{eqnarray} \label{ubs}
    U_{bs}|n,0\rangle = \sum_{p=0}^{n}{n\choose
    p}^{1/2}T^pR^{(n-p)}|p,n-p\rangle,
\end{eqnarray}
where $T$ and $R$ are the complex transition and reflection
coefficients, with normalisation $|T|^2 + |R|^2 = 1$. In general,
these resultant states will be entangled; indeed, for any input
made up of a finite superposition of Fock states the entanglement
cannot be zero (except, of course, in the case of the trivial
ground state). This can easily be proved by tracing one of the
output modes and checking that the resulting state (of the other
mode) is strictly mixed, i.e., its trace is less than $1$. Since
we will deal only with overall pure states in our paper, this
method for quantifying entanglement will be adequate in general
and we will use it later on in this section.

A Glauber state is defined as
\begin{eqnarray}
|\alpha\rangle =
\exp\left(-\frac{1}{2}|\alpha|^{2}\right)\sum_{n=0}^{\infty}\frac{\alpha
^{n}}{(n!)^{1/2}}|n\rangle
\end{eqnarray}
\cite{Glauber63}. When this enters one mode of a beam splitter and
a ground state the other, the output state is
\begin{eqnarray}
&&|\psi_{out}\rangle = U{bs}|\alpha,0\rangle \nonumber\\
&& =
\exp\left(-\frac{1}{2}|\alpha|^{2}\right)\sum_{n=0}^{\infty}\sum_{p=0}^{n}{n\choose
p}^{1/2}T^pR^{(n-p)}\frac{\alpha
^{n}}{(n!)^{1/2}}|p,n-p\rangle,\nonumber\\
\end{eqnarray}
which reduces to a product state
\begin{eqnarray}
|\psi_{out}\rangle =|\alpha/\sqrt(2)\rangle
\otimes|\alpha/\sqrt(2)\rangle.
\end{eqnarray}
It has been proven that the Glauber states are the only pure
states that when passed through one arm of a beam splitter return
product states with zero entanglement \cite{Aharonov66}. Therefore
they are the only classical states within this framework. Although
the beam splitter needs to be defined in a more general way (by
giving the transformation of the general basis state
$|n,m\rangle$, or, equally well, by defining the action of
annihilation and creation operators on different modes as in, for
example, \cite{Compos89}), it will for us be sufficient to use Eq.
(\ref{ubs}) as a specialised definition on the state
$|n,0\rangle$. We now ask what effect such a beam splitter has on
a spin-coherent state and to what extent the output state depends
on the magnitude of the spin.

A spin-coherent state $|z\rangle$ is defined here to be the
complex rotation of the ground state, parameterised by a complex
amplitude $z$ {\cite {Klauder,Radcliffe71,Arecchi72}}.--- In terms
of the spin raising operator $\hat{S}_+$ acting on the ground
state, we have that
\begin{eqnarray} \label{spin coh def1}
    |z\rangle = \frac{1}{(1+|z|^2)^S}\exp(\bar{z}\hat{S}_+/\hbar)|0\rangle.
\end{eqnarray}
Expanding the exponential we get
\begin{eqnarray} \label{spin coh def2}
|z\rangle &=& \frac{1}{(1+|z|^2)^S}\sum_{n=0}^{2S}{2S\choose
n}^{1/2}\bar{z}^n|n\rangle.
\end{eqnarray}
The summation only goes up to $2S$ since for $m = 2S$,
$\hat{S}_+|m\rangle = 0$. This can be seen easily using the
Holstein-Primakoff representation of spin operators in terms of
single-mode creation and annihilation operators \cite{Holstein40}.
The uncertainty relation for the spin operators, defined with
 the algebra $[\hat{S}_i,\hat{S}_j] = i \hbar \hat{S}_k$, is given by
\begin{eqnarray} \label{eq:uncertainty}
    \langle \hat{S}_i^2 \rangle \langle \hat{S}_j^2 \rangle \geq \frac{1}{4}\hbar^2 \langle \hat{S}_k \rangle ^2 .
\end{eqnarray}
This equality holds for the state $|z\rangle$; hence the
spin-coherent states are minimum-uncertainty states.

In \cite{Radcliffe71} and \cite{Arecchi72} it is shown by means of
the second quantisation, in terms of the operators generating the
state, that the limit takes the spin-coherent states to the
Glauber states. We now look at this limit purely in terms of the
first quantisation, which has the advantage that it allows us to
easily generate an infinite set of states that have the same
property that they asymptotically approach the Glauber states.
With the appropriate amplitude relationship, as $S$ goes to
infinity the spin-coherent states are equivalent to a limit of
Glauber states, namely, the infinite amplitude limit. Setting
$\alpha = \sqrt{2S}\bar{z}$ provides a suitable substitution. To
prove the equivalence it is enough to show that the overlap
between the Glauber state $|\alpha\rangle$ and the spin-coherent
state $|z'\rangle$ with this substitution goes to $1$ in the
infinite $S$ limit. That is, we wish to prove

\begin {eqnarray} \label{spinoptext}
  \lim_{S \rightarrow \infty} \langle \alpha|z'\rangle  &=&
    \lim_{S\rightarrow\infty}\left\{\exp{\left(-\frac{1}{2}|\alpha|^{2}\right)}\frac{1}{(1+\frac{|\alpha|^2}{2S})^S} \right. \nonumber\\
    && \left. \times\sum_{n=0}^{2S}\left(\frac{2S!}{(2S-n)!(2S)^n}\right)^{1/2}\frac{|\alpha|^{2n}}{n!}\right\}\nonumber\\
&=& 1 .
\end{eqnarray}

We can see that the normalisation term outside the sum clearly
goes to the exponential in the limit. We also see that each term
in the sum goes to $|\alpha| ^{2n}/n!$ in the limit. The limit of
the sum then gives an inverse exponential. Since both terms
converge in the same limit, the product of these terms converges
to the product of the convergences and so gives us a product of
two exponentials in the limit which cancel to give $1$. In fact,
this is true of any state $|\psi\rangle =
\sum_{n=0}^{d-1}f(n,d)[A^n/(n!)^{1/2}]|n\rangle$ such that $f(n,d)
> 0, \forall n,d \in \mathbb{N}$, and $f(n,d)$ converges to $1$ in
$n$ for all $p$. There are infinitely many such functions and
therefore these states define an infinite set of states that
asymptotically approach the Glauber state. A detailed proof can be
found in Appendix \ref{sc:appendix} (the orthogonality of these
states is discussed later). This result shows that, in the
infinite $S$ limit, the spin-coherent states do not entangle when
put through one arm of a beam splitter (since they are in fact
equivalent to Glauber states).

\begin{figure}[t]
\rotatebox{0}{\resizebox{!}{5cm}{\includegraphics{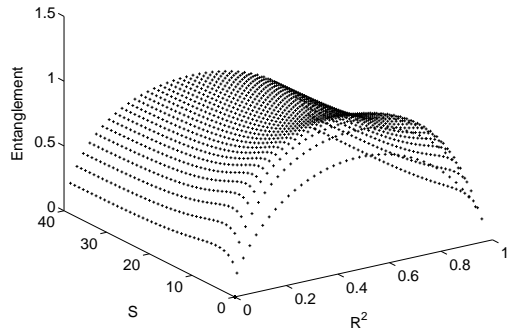}}}
\caption{\label{VNRSE}Entanglement of the
 state $|z,0\rangle$ after being passed through a beam splitter, against $S$ and $|R|^2$
for $|z|=3$. The maximum entanglement is given when $|R|^2 = 1/2$,
for all $S$.}
\end{figure}

We now wish to see exactly how the entanglement reaches this zero
limit. More precisely, we would like to determine if entanglement
falls off quickly with increasing spin, such that any reasonably
large spin system effectively returns a product state or if, for
instance, the drop-off is monotonic. The action of a beam splitter
on an input state, comprised of a spin-coherent state in one
mode/arm and the vacuum state in the other, gives
\begin{eqnarray} \label{ubsc}
    U_{bs}|z,0\rangle =
    \frac{1}{(1+|z|^2)^S}&&\sum_{n=0}^{2S}\sum_{p=0}^{n}\left\{{2S\choose
    n}^{1/2}{n\choose p}^{1/2}\right.\nonumber\\
    && \left.   \times T^pR^{(n-p)}\bar{z}^n|p,n-p\rangle\right\}.
\end{eqnarray}
We use the von Neumann entropy \cite{VonNeumann} as our measure
for entanglement, defined for a bipartite state $\rho_{AB}$ as
\begin{eqnarray}
\label{nent}
    E = -Tr{(\rho_A\ln\rho_A)} = - \sum_{i=0}^{i=2S}r_i\ln r_i,
\end{eqnarray}
where $\rho_A=Tr_B(\rho_{AB})$ is the reduced density matrix of
system $A$, and $r_i$ are its eigenvalues (and the squared Schmidt
coefficients of a pure state). In our case, from Eq. (\ref{ubsc}),

\begin{eqnarray} \label{reddm}
    \rho_A &=& \frac{2S!}{(1+|z|^2)^{2S}}\sum_{p,p'=0}^{2S}\sum_{m=0}^{\min(2S-p,2S-p')}\nonumber\\
   && \left\{\left(\frac{1}{p!p'!(2S-m-p)!
(2S-m-p')!}\right)^{1/2}\right.\nonumber\\
&&\left.\times\frac{|z|^{2m}}{m!}|T|^{2m}R^p\bar{R}^{p'}\bar{z}^pz^{p'}
|p\rangle\langle p'| \right\}.
\end{eqnarray}

We notice that in forming the reduced density matrix the complex
phases of $R$ and $z$ are effective only up to a unitary change of
basis, which does not effect its eigenvalues, and hence does not
affect the entanglement. Also, $\rho_A$ depends on only the
modulus of $T$. Thus from here on we are concerned only with the
modulus of these values: $|z|$; $|R|$ and $|T|$. In Fig.
\ref{VNRSE} we plot the entanglement against $S$ and $|R|^2$ for
$|z|=3$. For all $S$ we see a maximum entanglement for $|R|^2 =
1/2$. In fact, we found this for all $|z|$ checked between $0$ and
$50$. In addition, we can prove that setting $|R|=|T|= 1/\sqrt{2}$
gives the minimum linear entropy (see Appendix
\ref{sc:appendix2}), which is an upper bound to the von Neumann
entropy of entanglement, and can itself be regarded as a measure
of entanglement (see, for example, \cite{Bose00}). Since it is
analytically proven for the linear entropy, and it is confirmed by
all our numerical evidence, and the von Neumann and linear entropy
follow the same patterns in all our numerics, it is reasonable to
suppose that $|R|=|T|=\sqrt{1/2}$ does indeed give maximum von
Neumann entanglement. Hence from here on we take this to be the
appropriate value.

\begin{figure}[t]
\rotatebox{0}{\resizebox{!}{5cm}{\includegraphics{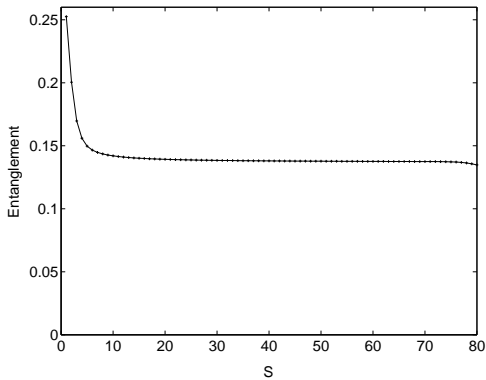}}}
\caption{\label{VNSE1}Entanglement of the
 state $|z,0\rangle$ after being passed through a beam splitter, against $S$
for $|z| = 1$. As we can see, the entanglement reduces quickly at
first but then more slowly. Indeed, it does not appear straight
away to limit to zero (al*though we know it does).}
\end{figure}

\begin{figure}[t]
 \rotatebox{0}{\resizebox{!}{5cm}{\includegraphics{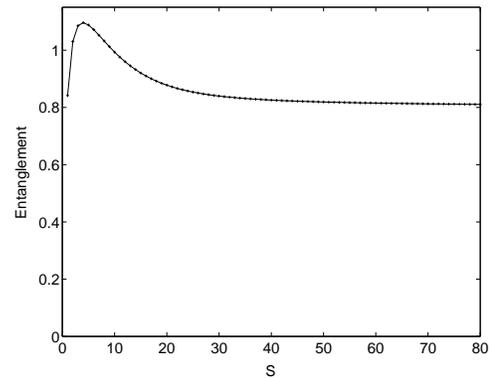}}}
 \caption{\label{VNSE3}Entanglement of the
 state $|z,0\rangle$ after being passed through a beam splitter, against $S$ for $|z| =
 3$. Here the entanglement rises first and then tails off after around
 $S=40$.}
 \end{figure}

\begin{figure}[hb]
\rotatebox{0}{\resizebox{!}{5cm}{\includegraphics{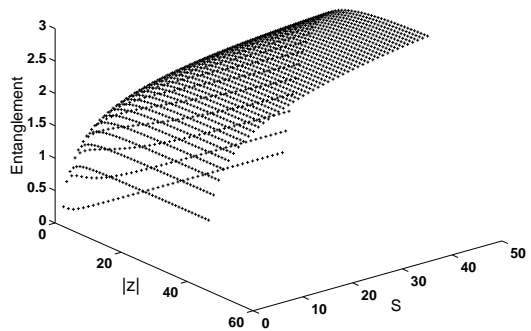}}}
\caption{\label{VNSEZ}Entanglement of the
 state $|z,0\rangle$ after being passed through a beam splitter, against $S$
and $|z|^2$. We see the pattern of a rapid initial peak followed
by very slow decrease thereafter for all $|z|$, and that an
increase in $|z|$ increases the entanglement for any $S$.}
\end{figure}

Two interesting features appear. The first is that the
entanglement does not quickly go to zero with increasing $S$, in
fact, after an initial quick change with amplitude $|z|$, the
entanglement settles to a very slow decline. The point where it
settles to the slow decline is very dependent on $|z|$. For higher
$|z|$ this point is both at a higher entanglement and for a higher
value of $S$. For each value of $|z|$ we see first a quick and
then a very slow fall from the maximum entanglement with $S$.  For
$|z|\leq1$ the maximum is at the origin, for higher $|z|$ we see a
peak, then the slow descent. For example, in Fig. \ref{VNSE1} we
see that the entanglement falls quickly to about $E=0.15$ at
around $S=15$, and then decreases very slowly thereafter (note
that, although it may not appear to, the entanglement does drop
with increasing $S$ - each consecutive point after $S=15$ is lower
than the previous one; the decline is just very slow).  In
contrast, in Fig. \ref{VNSE3} for $|z|=3$, we see a slow tailing
off from around $S=40$ at around $E=0.8$. In Fig. \ref{VNSEZ} we
plot the entanglement against $S$, from $S= 1/2 \rightarrow 50$
for $|z| =0 \rightarrow 50$. We can see that the peak where it
reaches maximum is higher for higher $|z|$. In this range of $S$,
the peak is not reached for $|z|$ larger than around $7$,
indicating that for such $|z|$ the drop to zero is even slower and
longer. Also, for any one $S$, the entanglement is higher for
larger $|z|$. It looks as if the entanglement settles down to a
value, dependent on $|z|$, from which it falls very slowly with
increasing $S$. If the rate of decline continues as in our
results, as we might expect, it would require huge $S$ to see the
zero limit approached. It seems the dependency on $|z|$ is far
greater than $S$.

The second feature of interest is that, for values of $|z|$ larger
than unity, the entanglement initially increases with increasing
$S$ (Figs. \ref{VNSE3} and \ref{VNSEZ}) and we see a peak. Since
we know that the entanglement must come arbitrarily close to zero
in the approach to the limit, we know it must go down from the
beginning value at $S=1/2$, and we might have expected this to be
a smooth fall.

The initial peak in generated entanglement, for $|z|$ larger than
around $1$, is best explained by looking at the output,
superposition state \ref{ubsc}. Roughly speaking, as we increase
$S$, the addition of entangled states to the superposition causes
a growth in entanglement. This is then countered as more states
are added, having the effect of diluting the superposition, so
that the entanglement goes down for greater $S$. We do not see
this for $|z|$ less than around $1$, because the coefficients for
the higher entangled states in the superposition decrease for
higher orders. We note that a similar phenomenon was seen in
\cite{Arnesen01}, where Arnesen {\it{et al}}. studied the
entanglement between two spin $1/2$ particles in a mixed thermal
state with an external magnetic field, according to the
antiferromagnetic Heisenberg model. For some field strengths, as
the temperature of the chain increased, there was an initial
increase in entanglement, although in the large limit the
entanglement went down to zero, as we would expect for such a
mixed system. The increase can be put down to the inclusion of
highly entangled states in the mixture, which then becomes
countered as the mixture dilutes as it extends to more states.
There is also a difference here in that in our case the increase
came from the addition of states to a pure superposition and in
theirs they are added to the mixture.

In addition to the subject of classicality discussed in the next
section, these results may be interesting for other reasons.
Coherent states are useful objects, with simple and continuous
parametrisation, they have many theoretical applications and allow
many simplifications (see \cite{Klauder} for many examples). They
can be created, for example, by a rotation of the ground state of
$N$ atoms \cite{Hepp73} or, in terms of the Schwinger spin states,
by optical fields passing through a beam splitter (e.g.,
\cite{Buzek89}). The latter can be squeezed and used to improve
interferometry \cite{Kitagawa93}. Beam splitters are also useful;
they are among the simplest devices implementable by experiment
and have many applications, including entanglement generation
itself (for example, \cite{Tan91}); so it is valuable to
understand them better in any scenario, especially one involving
entanglement generation, like this one. From this point of view it
might be interesting to note, for instance, that for an input
state with spin $S$, there exists an optimal amplitude $|z|$ that
returns the greatest entanglement.

It is also informative to consider the physical notion of a beam
splitter in the sense described above. In the case of an optical
state, when we use the term ``beam" we usually mean literally a
beam, or mode of light, that is defined by the path it takes. The
action of the beam splitter is then to split the incoming photons
of one mode (or path) into two, so the photons leave in a
superposition of both paths. This can be written in terms of
creation and annihilation operators, whence the beam splitter
annihilates photons from the input path and creates photons in
both output paths. With spin-coherent states this is not so clear.
The analogous operators to the annihilation and creation operators
are the step up and down operators. The analogy of creating a
photon in a beam would be to add a unit of spin to the system
(although the analogy is not exact because of different
commutation relations, which change the beam splitter
transformation, but broadly speaking it still accomplishes the
same). With this in mind one possible interpretation of a spin $S$
system is as a symmetric superposition of $2S$ spin $1/2$ systems.
Hence one can think of a beam of spin $1/2$ particles being split
into two paths, where the increase of one unit of spin is
equivalent to the addition of a spin $1/2$ particle. In this way,
the addition of more particles increases the total spin and so
makes the system more ``classical", as is the case for optical
states, where the more photons there are (the higher the
amplitude) the more classical the system can be said to be in
terms of distinguishability and entanglement (the states out of a
beam splitter are only the same as the input in the infinite limit
of $\alpha$). This view of a spin $S$ system has a natural link to
the Majorana representation to be discussed later, since
it can be seen as an illustration of these particles.\\

\bigskip
\section{Comments on Classicality} \label{sc:disc}

We now consider what the results of the previous section could
mean in terms of classicality. For the spin-coherent state, we do
not see the classical feature of the Glauber state of zero
entanglement generation when passed through one arm of a beam
splitter, except for those with the highest spin. Indeed, the
validity of the infinite limit case depends on the physical
situation at hand. It is possible though to imagine when such a
limit may be reasonable, for example, for a group of $10^5$ atoms
in a Bose-Einstein condensate. For the range we explored, $S=1/2
\rightarrow 50$, the entanglement is high and very dependent on
$|z|$. It is to be expected that this carries on for larger $S$
too. The infinite $S$ limit offers the set of ``most" classical
spin-coherent states, since they are equivalent to the Glauber
states and hence generate no entanglement. On this ground alone
though, the spin-coherent states are no more classical than
infinitely many other possible sets of states.

Then, the fact that we get an increase in entanglement as we
increase $S$ is surprising, in that it indicates a reduction in
classicality in this sense. Given this, we may ask if we can think
of a state that is more classical in this way; for example, one
that generates less entanglement and for which the entanglement
decreases more rapidly with the dimension of the Hilbert space.
This is a complicated problem since the only finite dimensional
state to give nonentangled states through a beam splitter is the
ground or vacuum state. The spin-coherent states can be
arbitrarily close to this zero entanglement, but that is simply
because they are arbitrarily close to the ground state in some
sense. We then ask how valid this measure of classicality is. For
example, we could think of other unitary transformations that
would create more entanglement with the input states here; indeed,
the amount of entanglement generated for a given transformation is
dependent on the input state. We can talk of the entangling
capacity of a unitary transformation \cite{Leifer02}, which is a
maximum taken over all input states; however, to define a class of
minimum entangling states for one unitary transformation does not
mean it is the class of minimum entangling states for a different
unitary transformation. The beam splitter transformation is a very
specific example, which is important in showing one classical
feature of the Glauber states, but one must ask if there is a
valid reason to take this one over other transformations as a
general measure. As such, we may claim that this is not an ideal
measure; indeed, as a concept, the resistance to entanglement
generation of a state is not well defined.

We can also consider the transition $S=1/2 \rightarrow \infty$
from other points of view. A paper by Lieb in 1973 showed that in
respect to the free energies of states, the spin states converge
to the classical case in the thermodynamic limit of a large number
and large $S$ spin states \cite{Lieb73}. This limit is very
similar to the one we take; however, here Lieb talks about a
thermal states, i.e., mixtures of spin-coherent states, whereas we
talk only about pure states. Other discussions of classicality
focus on the picture of the phase space, and how this can be said
to become that of a classical state (e.g., \cite{Hall02}). Indeed,
the use of coherent states to construct and compare phase spaces
can be used in many different situations, including the limit of
Lieb, as was discussed in detail by Yaffe in 1982 \cite{Yaffe82}.

We can look at the natural curvature of the phase space of the set
of spin-coherent states arrising from the Fubini-Study metric
\cite{Provost80}. The Riemannian curvature $K$, given by $K=1/2S$
is dependent on the size of the spin system. This can be seen by
imagining the phase gained by following a closed loop on the
surface of the sphere \cite{Vedral02}. In general, as a vector
follows a closed loop on a given space (maintaining parallel
transport), it gains a phase. This phase depends on the curvature
of the space; for a flat space no phase is acquired. As a
spin-coherent state follows a closed loop on the sphere, it will
gain a phase $\gamma = \oint \langle \psi (s)|d/ds|\psi (s)\rangle
ds$, which is equal to the curvature (up to a constant), and we
get $K=1/2S$. We see that for the ``least" classical, $S=1/2$,
state the curvature is maximum at $K=1$. This reduces smoothly and
monotonically with increasing $S$ until in the infinite limit, for
the ``most" classical state, the curvature is zero and that of the
Glauber states. This is also related to the distinguishability of
states, since the distance between two states (used in the
definition of this metric) increases as the size of the sphere
increases. Any two points on the phase space become further apart
(i.e., their overlap decreases). In this sense, the
distinguishability increases. Since in classical physics objects
should be distinguishable, in principle at least, this indicates
an increase in classicality. We note, though, that
distinguishability alone is not enough to claim classicality, for
example the Fock states, which exhibit clearly nonclassical
features, are orthogonal. However, it is an important feature of
classical physics and one that should be mentioned in any
discussion on classicality.

 We now look at the change in distinguishability through the transition $S=1/2 \rightarrow
\infty$ for two spin-coherent states, via the Helstrom optimal
measurement \cite{Helstrom}. We illustrate this change using the
Majorana representation \cite{Majorana32}.

One of the fundamental principles of quantum mechanics is the
uncertainty relation between incompatible observables which, in
turn, prohibits us from distinguishing two nonorthogonal states
perfectly with one measurement. This strongly contrasts with the
classical world where again, at least in principle, all states can
be distinguished perfectly. We can, however, optimise our
measurement strategy in quantum mechanics to give us the most
reliable answer possible. One such strategy is that given by
Helstrom \cite{Helstrom} (see also \cite{Fuchsthesis} for a very
readable account) where a measurement is made to give two
outcomes, one corresponding to each of the states, with a minimal
probability of error. We briefly state some of those results
before applying them to analysing spin-coherent states.

\begin{figure}[t]
\rotatebox{270}{\resizebox{!}{8.7cm}{\includegraphics{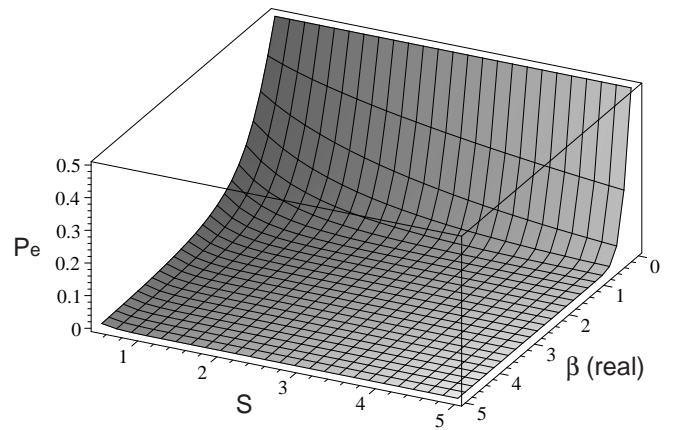}}}
\caption{\label{ProbError}Probability of error of distinguishing
two spin-coherent states $|0\rangle$ and $|\beta\rangle$ with
{\it{a priori}} probabilities $p_A$ and $p_B=1-p_A$, respectively,
against $S$ and $\beta (real)$ for $p_A=0.5$. The likelihood of
error quickly decreases with increasing $S$ for all $\beta$.}
\end{figure}

Suppose that we are given a system we know to be in one of two
states, $\rho_A$ or $\rho_B$, with probabilities $p_A$ and $p_B$,
respectively. When trying to ascertain which of the two
nonorthogonal states we have, we construct a two-outcome
generalised measurement, or positive operator-valued measure
(POVM), made up of $\hat{E}_A$ and $\hat{E}_B =
\hat{\one}-\hat{E}_A$, and associate the states with the
corresponding results. The probability of being incorrect in our
association, $P_e$, is given by
\begin{eqnarray}\label{PE1}
P_{e} = p_{A} \hbox{tr} (  \hat{E}_{B} \rho_{A}) + p_{B}\hbox{tr}
(\hat{E}_{A} \rho_{B}).
\end{eqnarray}
For us it is important that in the case of pure states $|A\rangle$
and $|B\rangle$ this POVM becomes a projection measurement with
the probability of error given by
\begin{eqnarray} \label{PE2}
P_{e}   &=& p_{A} \hbox{tr} (\hat{E}_B|A\rangle \langle A|) +
p_{B} \hbox{tr} (\hat{E}_A|B\rangle \langle B|).
\end{eqnarray}
Optimisation of this expression comes through our choice in the
projection measurements. Heuristically, we wish to choose the
projection spaces of $\hat{E}_A$ and $\hat{E}_B$ such that
$|A\rangle$ and $\hat{E}_B$ are as close to orthogonal as
possible; similarly for $|B\rangle$. This must be balanced against
the {\it{a priori}} probabilities to give the minimum probability
of error. When attempting to distinguish two spin-coherent states
$|\alpha\rangle$ and $|\beta\rangle$ we get
\begin{eqnarray} \label{PE3}
P_e = \frac{1}{2}\left(1-\sqrt{1-4p_Ap_B\left|\frac{(1+\alpha
\bar{\beta})^{2S}}{(1+|\alpha|^2)^S(1+|\beta|^2)^S}
\right|^2}\right).\nonumber\\
\end{eqnarray}

As mentioned earlier, as $S\rightarrow\infty$, all spin coherent
states become ``orthogonal'', that is, the overlap tends to zero.
This can be easily seen. Two spin-coherent states $|\alpha\rangle$
and $|\beta\rangle$ have overlap $| \langle \alpha| \beta \rangle
| =
|(1+\alpha\bar{\beta})^{2S}|/|(1+|\alpha|^2)^S(1+|\beta|^2)^S|$.
Since $|(1+\alpha\beta^*)^{2}|\leq|(1+|\alpha|^2)(1+|\beta|^2)|$,
the limit in $S$ gives $\lim_{S\rightarrow\infty } | \langle
\alpha | \beta \rangle | = 0$ except when the equality is reached,
i.e., $\alpha$ equals $\beta$, which, for all $S$, gives $ |
\langle \alpha | \alpha \rangle | = 1$. Thus, in the limit of
large $S$ all spin-coherent states become distinguishable and
$P_e$ becomes zero, (see Fig. \ref{ProbError}).

One very illuminating way to see the difference in the measurement
process as we change $S$ or $p_A$/$p_B$ is to represent the states
and the POVMs on the Riemann sphere using the Majorana
representation \cite{Majorana32} - which was also used by Zimba
and Penrose to recast the Kochen-Specker paradox \cite{Zimba93}.
In this system, a state with spin $S$ is represented by $2S$
points on a Riemann sphere. The position of these points is given
by the zeros of a complex function, constructed from the overlap
of the state with a non-normalised spin-coherent state. The
overlap between a non-normalised spin-coherent state
$|\tilde{z}\rangle$ and $|n\rangle$, is given by
\begin {eqnarray}
\langle \tilde{z}|n\rangle = {2S \choose n}^{1/2} z^n.
\end {eqnarray}
The general state $|\psi\rangle = \sum_{n=0}^{2S}a_{n}|n\rangle$
is described (up to a global phase) by the overlap function in $z$
(also known as the ``amplitude function" {\cite {Arecchi72}} or
the ``coherent-state decomposition'' {\cite {Leboeuf91}})
\begin{eqnarray} \label{function}
    \psi(z) = \langle \tilde{z} |\psi \rangle &=& \sum_{n=0}^{2s}{2s \choose n}a_{n}z^{n},\\
        &=& N \Pi_{i=1}^{2s}(z-z_{i}).
\end{eqnarray}
where $N$ is simply a normalisation constant.

The zeros of this function describe the state $|\psi\rangle$
entirely, again up to a global phase. To plot these points onto
the sphere, we view them as stereographic projections from the
north pole onto the complex plane going through the equator. Thus
a zero on the plane marks the south pole and as the modulus of the
root tends to infinity we get a point at the north pole - this
happens when $\psi(z)$ is of order less than the dimension $d$
minus $1$. The multiplicity of points at the north pole is equal
to the difference between the order of $\psi(z)$ and $d-1$; thus
the state $|0\rangle$ is represented by $2S$ points at the north
pole. The projection of $z = e^{i(\pi - \phi)}/\tan{(\theta/2)}$
onto the sphere represents the rotation of a point at the north
pole through angle $\theta$ around the axis $\hat{n} =
(\sin(\phi), - \cos(\phi),0)$, i.e., a point with azimuthal and
polar angles $\theta$ and $\phi$, respectively. This can also be
thought of as representing the rotation $\hat{R}_{\theta,\phi}$ on
state $|0\rangle$.

In the $S = 1/2$ case we find the well known Bloch sphere. Then,
the state $|\psi\rangle = \cos{(\theta /2)}|0\rangle +
e^{i\phi}\sin{(\theta/2)}|1\rangle$ has an overlap function
$\psi(z) = \cos{(\theta/2)} + e^{i\phi}\sin{(\theta/2)}$ with one
zero at $z = e^{i(\pi - \phi)}\tan{(\theta/2)}$, where $\theta$
and $\phi$ are the polar and azimuthal angles, respectively. We
also note that the Majorana picture could equally represent a
multiparticle state of $2S$ spin $1/2$ particles in a symmetric
superposition. Each point represents one spin $1/2$ particle
\cite{PenroseRindler}. Going back to the notion of a beam of spin
$S=1/2$ particles mentioned in the previous section, each point
would then represent a particle in our beam.

We can make a number of observations from this definition. First,
spin-coherent states are represented by $2S$ points all at the
same position on the unit two-sphere. This is clear since for a
spin-coherent state $|\alpha\rangle$ the overlap function is
$\langle \tilde{z} |\alpha \rangle =
(1+z\bar{\alpha})^{2S}/(1+|\alpha|^2)^S$, which has $2S$ zeros at
$-1/\bar{\alpha}$. Thus, the Majorana plot of a spin-coherent
state can be seen as a representation of the rotation
$\hat{R}_{\theta,\phi}$ that defines the state \ref{spin coh
def1}. Second, any rotation of a state $|\psi\rangle$ rotates the
sphere, since when forming the overlap function [Eq.
(\ref{function})], we can equally consider the rotation on $\psi$
as a reverse rotation on each of the $|\tilde{z}\rangle$'s,
corresponding to the zeros, which is also a rotation of the points
around the sphere. Note, though, that such a rotation can still
add a phase that will not be seen on the Majorana sphere. This
provides the basis for the next two points.

The modulus of the overlap between any two spin-coherent states is
given entirely by the angle between their points on the sphere.
This is true since any two pairs of spin-coherent states, with the
same angle between them on the Majorana sphere, are only a unitary
rotation apart (up to a phase); since this preserves the overlap
up to a phase, any such pairs have equal overlap modulus. Any
state with one or more points antipodal to a coherent state is
orthogonal to that state. To see this it is sufficient to show it
for the case $|\alpha\rangle = |0\rangle$, since we can simply
rotate the sphere and maintain any overlap up to a phase. The set
of states orthogonal to $|0\rangle$ are given by
$|\alpha^{\perp}\rangle = \sum_{n=1}^{2S}a_n|n\rangle$. The
coherent state decomposition thus has $z=0$ as one of its zeros at
least; indeed, this can occur only if the state has no components
in $|0\rangle$. This gives a Majorana point at the south pole,
i.e., antipodal.

With this, we can begin to look at how the Helstrom discrimination
strategy for two spin-coherent states can be seen in this
representation. The states themselves are shown as two points on
the sphere. To represent the projections we show the two resultant
states after the measurement takes place, $|e_A\rangle$,
$|e_B\rangle$, corresponding to $\hat{E}_A$ and $\hat{E}_B$,
respectively. We can do this because the two states to be
distinguished will both collapse onto the same one of two
resultant states. This is because any projection we make will be
on the two-dimensional subspace of the two states, since to
project outside this space would not be optimal. Hence, they
project onto the same pair of states in this subspace.

The first thing we notice is that for any $S$ these states give
Majorana points that lie on a circle on the surface of the sphere.
This is, in fact, the case for any state that is a superposition
of two spin-coherent states. For any complex coefficients $a$ and
$b$, the state $|\psi\rangle = a|0\rangle + b|\beta\rangle$ gives
overlap function zeros that lie on a circle of radius $r =
\left|a/b\right|^{1/2S}(1+|\beta|^2)^{1/2}/|\beta|$, centred at
$-1/\bar{\beta}$ on the complex plane. The stereographic
projection preserves circles and angles \cite{Eves}; hence the
Majorana points lie on a circle also (since this is true for the
above superposition, it is true for any superposition of two
coherent states because we can always rotate the pair so that one
state is $|0\rangle$ and maintain the geometry). Furthermore, the
radius and position of these circles relate back to the overlap of
the states and their {\it{a priori}} probabilities, although the
relationships are complicated, and we will talk about trends
qualitatively only.

\begin{figure}[t]
\rotatebox{270}{\resizebox{!}{8.7cm}{\includegraphics{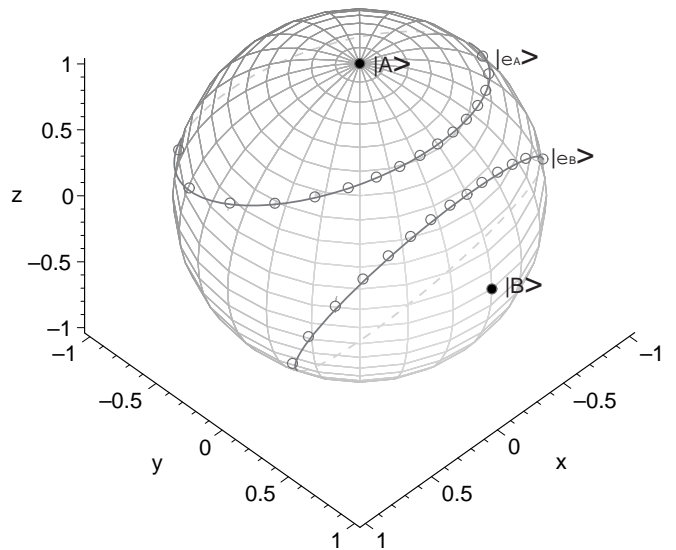}}}
\caption{\label{maj10}Majorana projection of two coherent states
$|A\rangle$ and $|B\rangle$, and the states postprojection
$|e_A\rangle$ and $|e_B\rangle$, for $S=10$ and $p_A = p_B =
0.5$.}
\end{figure}

In Fig. \ref{maj10} we see the Majorana projection of two
spin-coherent states, given by two black points, and the states
post projection, given by the grey points, which exist on two
circles, for $S=10$. State $|A\rangle = |0\rangle$ is on the north
pole and state $|B\rangle = |-i\rangle$ is at $(1,0,0)$. The
{\it{a priori}} probabilities are set at $p_A = p_B = 0.5$.

We can now look at how when changing the {\it{a priori}}
probabilities $p_A$/$p_B$ and $S$, the Helstrom measurements we
make, found from the minimisation of Eq. (\ref{PE2}), also change.
For any fixed spin, as the {\it{a priori}} probability of
$|A\rangle$ increases, the circle of the corresponding projection
$\hat{E}_A$ closes in around the state, until it becomes that
state. That is, $|e_A \rangle \rightarrow |A\rangle$, as we would
expect for minimisation of the error (\ref{PE2}). The other
circle, for projection $\hat{E}_{B}$, widens until it becomes the
state orthogonal to $|A\rangle$ in the $|A\rangle$,$|B\rangle$
subspace ($|A_{\perp}\rangle$), this gives the circle whose width
allows one point to exist antipodal to the Majorana point of
$|A\rangle$. The opposite to this occurs should $p_B$ increase.

Keeping $p_A, p_B$ fixed and increasing $S$, both circles become
larger, as $|e_A \rangle \rightarrow |B_\perp\rangle$ and $|e_B
\rangle \rightarrow |A_\perp\rangle$. This shows the reason for
the decrease in probability of error with $S$. As $S$ increases,
the states $|A\rangle$ and $|B\rangle$ become more orthogonal;
this allows the projection POVMs to change to become more
orthogonal to the anticorresponding states, in accordance with the
minimisation of Eq. (\ref{PE2}), seen by the growth of the
circles, where $P_e$ then reduces with increasing $S$. Therefore,
the smooth growth of the projection circles illustrates that,
within this distinguishability criterion for classicality,
spin-coherent states approach the ``classical", Glauber, states
with the increase of spin. This is in contrast with the previous
criterion based on entanglement generation where we see a brief
increase of entanglement with spin.

\section{Conclusion} \label{sc:conc}

In this paper, we have examined the transition of spin-coherent
states from spin $S=1/2$ to the limit $S \rightarrow \infty$ with
respect to the generation of entanglement through a beam splitter.
We gave an alternative proof that, as the spin tends to infinity,
the spin-coherent states tend to the infinite amplitude limit of
the Glauber states. From this proof we can easily see that this is
not unique to spin coherent states; indeed, there are infinitely
many sets of states for which this is true. The spin-coherent
state does represent an element of these that achieves minimum
uncertainty.

We have studied the entanglement generation of spin-coherent
states for the specific case where they are sent through one arm
of a 50:50 beam splitter and a vacuum through the other. Two main
features were observed. The first is that the entanglement of the
output states depends heavily on $|z|$, more than $S$. For all
$|z|$, after an initial quick change with $S$, the entanglement
settles to a very slow, almost flat decline. The entanglement at
which this slow tailing off begins is very dependent on $|z|$ and
is higher for larger $|z|$. To see the high $S$ near zero
entanglement, one would need a system of huge spin number. The
relevance of this limit is very dependent on the physical system.
In our numerics we considered only up to spin $S=50$, but we may
see a significantly low resultant entanglement for systems with
extremely large $S$, for example, of the order of $10^5$ spin
$1/2$ atoms. The second interesting feature is that for $|z|>1$
the entanglement initially rose with $S$.

We then discussed this transition in terms of classicality. The
high dependence on $|z|$, very slow decline to zero, and increase
of entanglement (thus dip in this sense of classicality) are
indicators that this measure may be inadequate as a universal
signature of classicality, as we found. The value in these results
is in the understanding of a beam splitter as an entanglement
generator, with respect to which it is also found that the
generated entanglement increases with amplitude $z$, and that for
any spin $S$ we can find an amplitude with $|z|$ that gives the
maximum entanglement at output. In a sense we have sacrificed the
generality of this result as a classicality measure for its
simplicity and applicability.

We listed briefly some examples of other ways the transition can
be looked at in terms of classicality. The large $S$ limit
discussed by Lieb also leads to a classical description, in terms
of free energies. The change in phase space as we change $S$ also
indicates a transition to classicality in some sense. We then
focused on distinguishability in terms of state discrimination
using the Helstrom optimal measurement strategy and illustrated it
using the Majorana picture. The geometry of the states and the
Helstrom POVMs is simple, and allows us to illustrate easily the
changes affecting the probability of error for the measurement. A
spin-coherent state is simply one point on the sphere and the
POVMs are represented by circles of points. The decrease in
probability of error with an increase in $S$ is seen by the
increase in the diameter of the circles on the sphere representing
the measurements. We reiterate that distinguishability does not
constitute classicality in its own right. The standard example of
Fock states makes is an example. Indeed none of these measures
give a common result; they all follow different declines and
cannot be said to be equivalent exactly in the transition regime.
Distinguishability is important and although not a complete
description, if we want to look at classicality, we must discuss
this, as we must all areas.

Possible extensions of this work would naturally include the move
to considering the mixed case as in \cite{Kim02} and
\cite{Xiang-bin02}. A consideration of a thermal mixture of
states, as in Lieb's calculation, may offer interesting results in
terms of entanglement. It would also be interesting for future
work to create common criteria for ``classicality measures" or
indicators of (pure) states. It seems that one common feature that
should be imposed is the increase in classicality with the size of
a system, or the number of subsystems involved, so that it appears
classical in the large macroscopic limit. An important problem,
brought to light in the discussion of the beam splitter, is the
assignment of classicality in terms of entanglement. A classical
system should be inefficient at generating entanglement within
itself, but on the other hand, should be able to strongly entangle
with its environment (this is required, for example, in
decoherence models). It is not at all clear that there exists a
single measure that would capture all these desirable properties
and this remains a challenging open problem.

\bigskip
\par {\em Acknowledgements.}
We would like to thank William Irvine, Manfred Lein, and Stefan
Scheel for very useful discussions on the subject of this paper
and Jens Eisert for critical reading of the proofs involved. This
work was sponsored by the Engineering and Physical Sciences
Research Council, the European Community, the Elsag spa and the
Hewlett-Packard company.

\appendix
\section {} \label{sc:appendix}
Glauber states are defined as \cite{Glauber63}
\begin {eqnarray} \label{OpCohState}
|\alpha \rangle &=& \exp\left(\alpha \hat{a}^{\dag} -
\frac{1}{2}|\alpha|^{2}\right)|0\rangle
\nonumber\\
&=&\exp\left(-\frac{1}{2}|\alpha|^{2}\right)\sum_{n=0}^{\infty}\frac{\alpha
^{n}}{(n!)^{1/2}}|n\rangle.
\end{eqnarray}
\\
We wish to show that, by appropriate substitution, the
spin-coherent state (\ref{spin coh def2}) is equal to a subclass
of Glauber states as $S$ tends to infinity. We know that for any
two different spin-coherent states, the overlap tends to zero as
$S$ tends to infinity; thus the ``amplitude" $\alpha$ for the
optical case must tend to infinity as $S$ does. We set $\alpha =
\sqrt{2S}\bar{z}$. Substituting this into Eq. (\ref{spin coh
def2}) we get
\begin {eqnarray} \label{spinop1}
    |z'\rangle &=& \frac{1}{(1+\frac{|\alpha|^2}{2S})^S}\
\sum_{n=0}^{2S}{2S\choose n}^{1/2}
\left(\frac{\alpha}{2S}\right)^{n}|n\rangle.
\end{eqnarray}
\\
To prove the equivalence of the two states in the infinite limit,
it is sufficient to show that the overlap of a Glauber state
$|\alpha\rangle$ and such a spin-coherent state goes to $1$ (i.e.,
they are the same state). Thus, we wish to prove that

\begin {eqnarray} \label{spinop}
  \lim_{S \rightarrow \infty} \langle \alpha|z'\rangle  &=&
    \lim_{S\rightarrow\infty}\left\{\exp{\left(-\frac{1}{2}|\alpha|^{2}\right)}\frac{1}{(1+\frac{|\alpha|^2}{2S})^S} \right. \nonumber\\
    && \left. \times\sum_{n=0}^{2S}\left(\frac{2S!}{(2S-n)!(2S)^n}\right)^{1/2}\frac{|\alpha|^{2n}}{n!}\right\}\nonumber\\
&=& 1 .
\end{eqnarray}

Let us first look at the summation.\\
{\it{Lemma}}. For all $A \in \mathbb{R}^+$,
\begin {eqnarray} \label{spinop2}
    \lim_{n\rightarrow\infty } \sum_{p=0}^{n}\left(\frac{n!}{(n-p)!n^{p}}\right)^{1/2}\frac{A^{p}}{p!}
    = \sum_{p=0}^{\infty}\frac{A^{p}}{p!}.
\end{eqnarray}
\\
{\it{Proof}}. We will use the method of showing that the upper and
lower bounds to $\langle \alpha | z'\rangle$ coincide as $S$ tends
to infinity. Since each term in the sum is positive, one upper
bound of the left hand side of Eg. (\ref{spinop2}) can be found by
taking the sum to infinity. Thus
\begin {eqnarray} \label{spinop3}
    \lim_{n\rightarrow\infty } \sum_{p=0}^{n}\left(\frac{n!}{(n-p)!n^{p}}\right)^{1/2}\frac{A^{p}}{p!} &\leq&  \lim_{n\rightarrow\infty }  \sum_{p=0}^{\infty}\left(\frac{n!}{(n-p)!n^{p}}\right)^{1/2}\frac{A^{p}}{p!}\nonumber\\
    &\leq& \sum_{p=0}^{\infty}\frac{A^{p}}{p!},\nonumber\\
    &&
\end{eqnarray}

where the last line is given by taking the limit inside the sum,
which we are allowed to do since each term converges.

We then find a lower bound by taking the sum to some finite $m$,
so
\begin {eqnarray} \label{spinop4}
    \lim_{n\rightarrow\infty } \sum_{p=0}^{n}\left(\frac{n!}{(n-p)!n^{p}}\right)^{1/2}\frac{A^{p}}{p!}
    &\geq& \lim_{n\rightarrow\infty }
    \sum_{p=0}^{m}\left(\frac{n!}{(n-p)!n^{p}}\right)^{1/2}\frac{A^{p}}{p!}.\nonumber\\
    &&
\end{eqnarray}

However, taking the infinite limit of $m$ this inequality still
holds, since it is true for any finite $m$ and so it is true for
any arbitrarily small distance from the limit, so the limit can at
most be equal. This gives the same upper as lower bound; hence

\begin {eqnarray} \label{spinop5}
    \lim_{n\rightarrow\infty } \sum_{p=0}^{n}\left(\frac{n!}{(n-p)!n^{p}}\right)^{1/2}\frac{A^{p}}{p!}
    &=& \lim_{n\rightarrow\infty }
    \sum_{p=0}^{\infty}\left(\frac{n!}{(n-p)!n^{p}}\right)^{1/2}\frac{A^{p}}{p!}\nonumber
    \\
    &=& \sum_{p=0}^{\infty}\frac{A^{p}}{p!}\nonumber \\
    &=& \exp(A).
\end{eqnarray}

Since the normalisation and the summation in Eq. (\ref{spinop})
converge with increasing $n$, the limit of the product is the
product of the limits and thus

\begin {eqnarray} \label{spinop6}
    \lim_{S \rightarrow \infty} \langle \alpha|z'\rangle  &=& \exp{\left(-|\alpha|^{2}\right)}\exp{\left(|\alpha|^{2}\right)}\nonumber
\\
&=& 1 .
\end{eqnarray}
Interestingly, this is also true of any state $|\psi\rangle =
\sum_{n=0}^{d-1}f(n,d)[A^n/(n!)^{1/2}]|n\rangle$ such that $f(n,d)
> 0, \forall n,d \in \mathbb{N}$, and $f(n,d)$ converges to $1$ in
$n$, for all $p$, for example, $|z\rangle =
\exp\left(-\frac{1}{2}|z|^{2}\right)\sum_{p=0}^{n}(z^p/p!)
|p\rangle$. There are infinitely many such functions and therefore
there are infinitely many states that will tend to the Glauber
state in the limit, as stated in Sec. \ref{sc:BS}.

\section {} \label{sc:appendix2}

Linear entropy is an upper bound to the von Neumann entropy and is
defined as
\begin{eqnarray}
    S_{lin} = 1-Tr\left(\rho_A^2\right).
\end{eqnarray}

The linear entropy of the two output ``beams" is given by
\begin{eqnarray}
    S_{lin} &=& 1-\frac{(2S!)^2}{(1+|z|^2)^{4S}}\sum_{p,p'=0}^{2S}\sum_{m,m'}^{min(2S-p,2S-p')}\left(\frac{|z|^{m+m'+p+p'}}{m!m'!}\right.\nonumber\\
    && \left. \times \frac{1}{\sqrt{(2S-m-p)!(2S-m-p')!}}\right.\nonumber\\
    && \left. \times \frac{1}{\sqrt{(2S-m'-p)!(2S-m'-p')}}\right.\nonumber\\
    & & \left.\times |R|^{2(p+p')}|T|^{2(m+m')}\right),
\end{eqnarray}

with $|R|^2+|T|^2=1$. Differentiating this with respect to $|R|$,
we have
\begin{eqnarray} \label{diff}
    \frac{dS_{lin}}{d|R|} &=&
    -\sum_{p,p'=0}^{2S}\sum_{m,m'}^{min(2S-p,2S-p')}\left\{f(|z|,S,m,m',p,p')\right. \nonumber\\
     &\times& \left. \left(|R|^{(p+p'-1)}(p+p')(1-|R|)^{(m+m')}\right. \right.\nonumber\\
    && \left.\left. -(1-|R|)^{(m+m'-1)}(m+m')|R|^{(p+p')}\right)\right\},
\end{eqnarray}
where
\begin{eqnarray}
    f(|z|,S,m,m',p,p') = \frac{(2S!)^2}{(1+|z|^2)^{4S}}\left(\frac{|z|^{m+m'+p+p'}}{m!m'!}\right.\nonumber\\
     \left. \times \frac{1}{\sqrt{(2S-m-p)!(2S-m-p')!}}\right.\nonumber\\
     \left. \times \frac{1}{\sqrt{(2S-m'-p)!(2S-m'-p')}}\right) .
\end{eqnarray}

Using the symmetry of the summations and the symmetry of
$f(|z|,S,m,m',p,p')$, one can see that Eq. (\ref{diff}) gives a
zero at $|R|=|T|=1/\sqrt{2}$. Hence the linear entropy is maximum,
for any $S$ and $z$, when $|R|=|T|=1/\sqrt{2}$.

\bibliographystyle{unsrt}
\bibliography{GenBib}

\end{document}